\documentclass{IEEEtran}
\usepackage{cite}
\usepackage{amsmath,amssymb,amsfonts}
\usepackage{graphicx}
\usepackage{textcomp,nicefrac}

\begin{document}

\title{Time resolution of BC422 plastic scintillator \\
read out by a SiPM}

\author{Alexey Stoykov and Tigran Rostomyan%


\thanks{Manuscript received March 22, 2021; accepted June 13, 2021.}
\thanks{\copyright 2021 IEEE.  Personal use of this material is permitted.  Permission from IEEE must be obtained for all other uses, in any current or future media, including reprinting/republishing this material for advertising or promotional purposes, creating new collective works, for resale or redistribution to servers or lists, or reuse of any copyrighted component of this work in other works.}
\thanks{A. Stoykov and T. Rostomyan are with the Paul Scherrer Institut, CH-5232 Villigen PSI, Switzerland (e-mail: alexey.stoykov@psi.ch)}}

\markboth{IEEE Transactions on Nuclear Science,~Vol., No.~, March~2021}%
{Shell \MakeLowercase{\textit{A. Stoykov et al.}}: Time resolution of BC422 plastic scintillator read out by a SiPM}

\maketitle

\begin{abstract}
Plastic scintillators are widely used in particle detectors when precise timing information is required. 
As a basis for comparing different detectors we use such a characteristic as the time resolution per 1\,MeV detected energy, i.e. the expected time resolution assuming the deposited energy in the scintillator equals to 1\,MeV and all the scintillation light is collected to the photosensor. In this work we measure this parameter with BC422 plastic scintillator read out by different SiPMs. The best obtained value is about 6\,ps, which is almost a factor of 3 better compared to our earlier result from 2012.  Such an improvement represents the progress in the development of SiPMs made over these years and, driven by this progress, an improved knowledge and know-how on such detectors.
\end{abstract}



\section{Introduction}

\IEEEPARstart{A}{t} the Paul Scherrer Institut (PSI) in Switzerland we routinely use plastic scintillators read out by silicon photomultipliers (SiPMs) to build detectors for various experiments \cite{Stoykov12a,Cattaneo14,Amato17,Rostomyan21}. 
The basis for estimating the time resolution of such detectors was put in the research works \cite{Stoykov12b,Sedlak12}. It was shown both experimentally and in simulations, that in case of using a constant fraction discriminator (CFD), the time resolution $\sigma$ is inversely proportional to the square root of the energy $E$ deposited in the scintillator. Here, we represent such dependence in the following form:

\begin{equation}
\sigma = \frac{\sigma_{\rm 1MeV}}{\sqrt{k E}}\ ,
\label{Eq:sigma=f(E)}
\end{equation}
where $k$ is the fraction of emitted scintillation light collected to the photosensor (single SiPM or group of SiPMs); $k E = E_{\rm det}$ is the energy effectively reaching the photosensor (detected energy); $\sigma_{\rm 1MeV}$, expressed in ${\rm ps} \times \rm{MeV}^{0.5}$, is a coefficient, which can be interpreted as the time resolution per 1\,MeV detected energy.

The best value of $\sigma_{\rm 1MeV}$, reported in \cite{Stoykov12b}, is 18\,ps$\times$MeV$^{0.5}$, assuming $k = 1$. At that time this was the first proof that with a SiPM based detector one can have as good time resolution as with using a PMT. Taking a more appropriate estimate $k \approx 0.74$  (see below) for the light collection under the given conditions, we get $\sigma_{\rm 1MeV} \approx 15\,{\rm ps} \times {\rm MeV}^{0.5}$. None of the detectors in \cite{Stoykov12a,Cattaneo14,Amato17,Rostomyan21} exceeded this performance limit. 

In \cite{Gundacker20} the time resolution with different scintillators and SiPMs was studied using a leading edge (LE) discriminator technique. A fair agreement between the experimental time resolution limit and predictions of an analytical formula \cite{Vinogradov18}, estimating an optimum time resolution, was found. 
Following \cite{Gundacker20,Vinogradov18}, the lower limit of the parameter $\sigma_{\rm 1MeV}$ in (\ref{Eq:sigma=f(E)}) in case of using a LE discriminator can be estimated as:

\begin{equation}
\sigma_{\rm 1MeV,min} = 
\sqrt{\frac{{\rm ENF} \times \tau_{\rm d} \times (1.57\,\tau_{\rm r} + 1.13\,\sigma_{\rm ts})}
{{\rm LY} \times {\rm PDE}} }\  ,
\label{Eq:sigma1MeVmin}
\end{equation}

where $\tau_{\rm r}$ and $\tau_{\rm d}$ are the rise and decay times of the scintillation emission
with condition $\tau_{\rm r} \ll \tau_{\rm d}$;
LY is the scintillation light yield per unit energy; 
PDE and ENF are the photon detection efficiency and the excess noise factor of the SiPM;
$\sigma_{\rm ts} = \sqrt{\sigma_{\rm PTS}^2 + \sigma_{\rm SPTR}^2}$ is the standard deviation characterizing cumulative time spread in the light collection and SiPM response formation processes, with abbreviations PTS and SPTR standing for the photon transfer time spread and single photon time resolution, respectively. 
In \cite{Gundacker20} the comparison of the experimental data with the model \cite{Vinogradov18} 
is done with ${\rm ENF} = 1$; the reasons for neglecting the excess noise contribution are, however, not provided.

Note, that if one takes the datasheet value $\tau_{\rm r} = 350$\,ps for the rise time of the scintillation emission of BC422 \cite{Bicron} (equivalent EJ-232 \cite{Eljen}), the timing properties of the SiPM are expected (\ref{Eq:sigma1MeVmin}) to have only a minor influence on the achievable time resolution, as for modern SiPMs the $\sigma_{\rm SPTR}$ values are below 100\,ps \cite{Gundacker20}.
However, in \cite{Gundacker20} the authors pointed out, that the scintillation rise time of BC422 is actually a factor of 10 lower, compared to its datasheet value. From this they naturally concluded that the time resolution of BC422 coupled to a SiPM is not limited by the scintillator emission properties and should be more influenced by the SPTR value of the SiPM. Their detector consisted of a small ($3 \times 3 \times 3$)\,mm$^3$ cube scintillator wrapped in Teflon tape attached to one SiPM (estimated light collection $k \approx 0.74$). The coincidence time resolution (CTR) between two such detectors for 511\,keV annihilation photons from a $^{22}$Na radioactive source was measured for a narrow energy window set around the Compton edge with the mean value of 340\,keV. A low-noise signal processing at 1.5\,GHz bandwidth with a low-threshold LE timing was used. The CTR value of 35\,ps (FWHM) or 15\,ps (sigma) was reported, which gives $\sigma_{\rm 1MeV} = 5.3\,{\rm ps} \times {\rm MeV}^{0.5}$.

This result shows a factor of three improvement compared to that of \cite{Stoykov12b}, indicating  
that our data on $\sigma_{\rm 1MeV}$ need updating. In this work, we perform measurements with BC422 scintillator and different SiPMs. We decided against repeating the signal-processing scheme of \cite{Gundacker20} and utilized our standard approach with signal bandwidth of $\sim 600$\,MHz and a CFD. 
The formula (\ref{Eq:sigma1MeVmin}) serves us only as a guidance on how different scintillator and SiPM parameters influence the time resolution; no verification of it with the current data is aimed for.

\section{Experimental Conditions}

\subsection{Scintillators and SiPMs}

\begin{figure}[!h]
\centering
\includegraphics[width=1.0\columnwidth,clip]{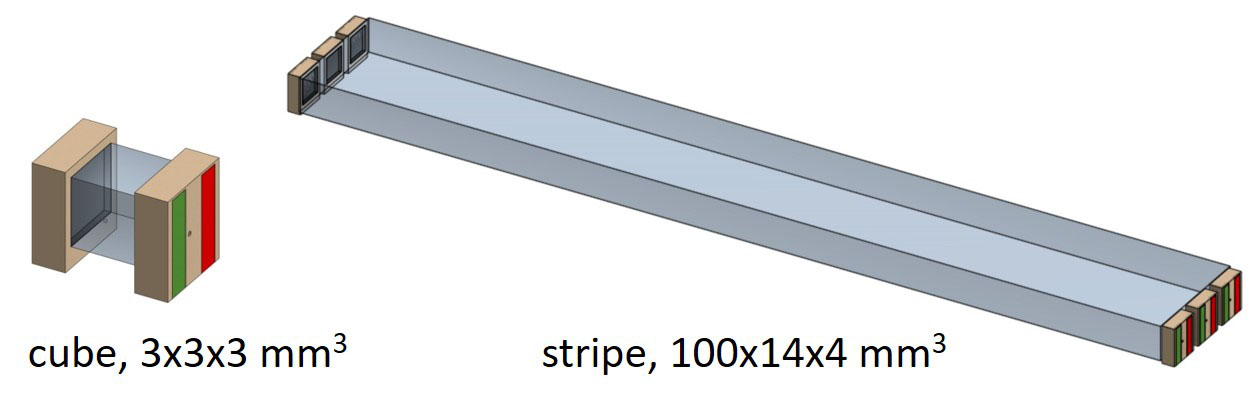}
\caption{
Scintillation detectors used in this work.
}
\label{Fig:scint}
\end{figure}

Most of the measurements in this work are done with ($3 \times 3 \times 3$)\,mm$^3$ cubes of BC422 or EJ-232 plastic scintillator (see Fig.\,\ref{Fig:scint}). Initially, all scintillator faces are polished with a diamond tool. Two SiPMs of the same type are attached to opposite faces of the cube using either optical cement EJ-500 or optical grease EJ-550. Three types of such detectors are used:

\begin{itemize}
\item{
(cube, Teflon, cement): scintillator faces being attached to SiPMs are rough grinded, coupling to SiPMs is done via optical cement, afterwards  the detector is wrapped in a Teflon tape;}
\item{
(cube, black, cement): all faces of the scintillator are rough grinded, the SiPMs are attached with optical cement, afterwards the ``free`` faces are painted with black paint EJ-510BL;}
\item{
(cube, black, grease): scintillator faces being attached to SiPMs are left polished, the ``free`` faces are rough grinded and painted with black paint, coupling to SiPMs is done via optical grease.}
\end{itemize}

The small cube detectors are used to obtain the best possible $\sigma_{\rm 1MeV}$ value for each SiPM type. However, one can ask up to which extent such results reflect the situation with larger detectors, used in practice. To answer this question, we also perform measurements with a relatively large detector built as a stripe ($100 \times 14 \times 4$)\,mm$^3$ of BC422 scintillator with 3 SiPMs coupled to each ($14 \times 4$)\,mm$^2$ face and electrically connected together in series 
(Fig.\,\ref{Fig:scint}). All scintillator faces are diamond polished, the coupling to SiPMs is done via optical grease.

The tested SiPMs are listed in Table~\ref{Table:SiPMs}. Two SiPMs with lowest SPTR, reported in \cite{Gundacker20}, are non-commercial devices and were not available for the current measurements.

\begin{table}[!htb]
\centering
\renewcommand{\arraystretch}{1.1}
\caption{
List of SiPMs used in this work. $U_0$  is the typical breakdown voltage and $\Delta U_1$  is the chosen operation overvoltage per 1 SiPM, also in case of their series connection. Hamamatsu S10931-050PX is an old device similar to the one used in \cite{Stoykov12b}. Active area of Broadcom SiPM is ($3.72 \times 3.72$)\,mm$^2$; for all others is ($3.0 \times 3.0$)\,mm$^2$
}
\begin{tabular}{clcc}
\hline
\lower 5pt \hbox{Producer}
  & \lower 5pt \hbox{SiPM}
  & \lower 5pt \hbox{$U_0$,\,V}
  & \lower 5pt \hbox{$\Delta U_1$,\,V} \\[10pt]
\hline
\phantom{00}Broadcom	& AFBR-S4N44C013 	& 27 		& \phantom{1}6.7 \\
\phantom{00}Advansid	& ASD NUV3S-P       	& 26 		& \phantom{1}3.5 \\
\phantom{00000}Ketek	& PM3325-WB-D0	 	& 25 		& \phantom{1}5.7 \\
\phantom{}Hamamatsu	& S10931-050PX		& 70		& \phantom{1}1.8 \\
\phantom{}Hamamatsu	& S12572-025P			& 65		& \phantom{1}4.2 \\
\phantom{}Hamamatsu	& S13360-3050PE		& 53		& \phantom{1}7.4 \\
\phantom{}Hamamatsu	& S14160-3015PS		& 38		& 10.0 \\
\phantom{}Hamamatsu	& S14160-3050HS		& 38		& \phantom{1}3.5 \\
\hline
\end{tabular}
\label{Table:SiPMs}
\end{table}

\subsection{Measurement Setup}

Figures\ 2 and 3 show two setups used to perform the measurements. The scheme of the experiment is the same in both cases and comprises a $^{90}$Sr radioactive source, a collimator, a detector under test D2, and a reference detector D1.  The collimator is a 1\,mm thick brass plate with $\varnothing 1.2$\,mm hole. The detector under test D2 is the one for which the construction (SiPMs) and the signal processing (amplifier and CFD settings) will be varied to find their impact on $\sigma_{\rm 1MeV}$. The construction of the reference detector D1 is chosen to ensure the best timing performance: BC422 cube wrapped in Teflon with Broadcom SiPMs coupled via optical cement.  After calibration, in which the time resolution of this detector is established, the corresponding signal processing parameters are fixed throughout the measurements.

\begin{figure}[!htb]
\centering
\includegraphics[width=1.0\columnwidth,clip]{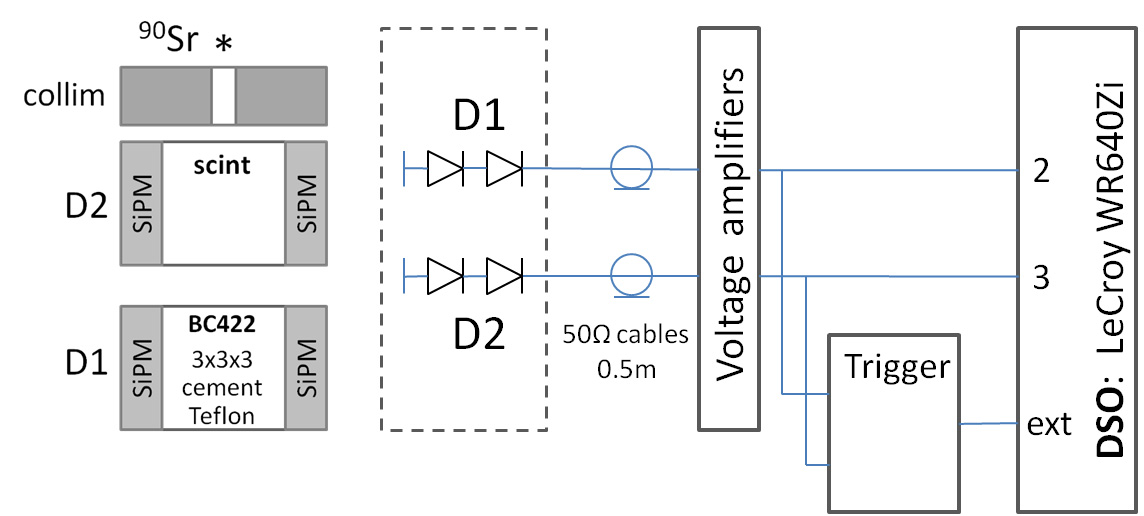}
\caption{
Measurement setup 1. Two SiPMs in each D1 and D2 detectors are connected in series and perform as a single photosensor. No compensation for the particle hit position is possible; some degradation of the time resolutions is to be expected.
}
\label{Fig:setup1}
\end{figure}

\begin{figure}[!htb]
\centering
\includegraphics[width=1.0\columnwidth,clip]{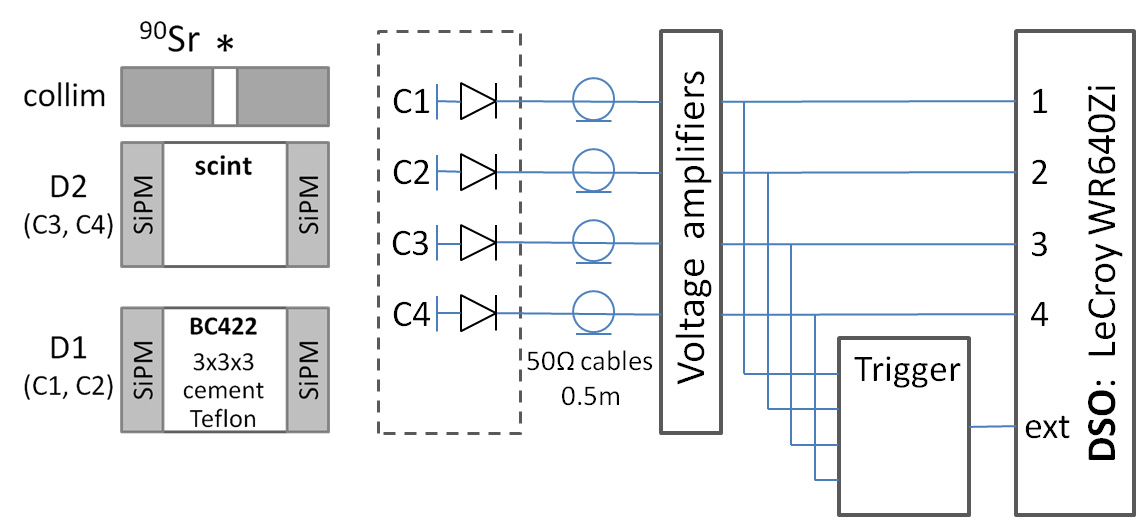}
\caption{
Measurement setup 2. Two SiPMs in each detector are readout independently which allows, by averaging, 
a compensation for the particle hit position.
}
\label{Fig:setup2}
\end{figure}

In setup 1 (Fig.\,\ref{Fig:setup1}) the two SiPMs in D1 and D2 are connected together acting as a single photosensor. We use a series connection of SiPMs both for DC and AC current. The signals D1 and D2 are first sent via 0.5\,m long 50\,Ohm cables to a voltage amplifier. After the amplifier each signal is split using a passive 50\,Ohm splitter between a trigger logic and a digital oscilloscope (DSO, model LeCroy WaveRunner 640Zi \cite{LeCroy}). The trigger logic is realized in NIM modules. Its function is to set a detection threshold for each channel and generate a trigger signal for the DSO. In calibration mode the trigger is generated as tr\ =\ (D1)D2, i.e. coincidence D1 and D2 taken on the leading edge of D1; in measurement mode tr\,=\,D1. The DSO is operated at the sampling rate of 40\,GS/s. The bandwidth for each channel is adjusted by tuning the ERes (enhanced resolution) filter of the DSO. This is a sort of a moving average filter trading bandwidth for vertical resolution, which can be enhanced up to 3\ bits in steps of 0.5\ bit. In all, except one, measurements the ERes bandwidth is set to $\sim 600$\,MHz. The DSO measures charge spectra of signals D1 and D2, and perform analog CFD transformation on them, generating signals F1 and F2. The detection times $t_1$ and $t_2$ are taken at the zero-crossing of F1 and F2. The collected histograms $t_2 - t_1$ are used in analysis to extract the CTR values. The electronic noise contribution (mainly DSO) to the time jitter of F1 and F2 at their zero-crossing is also measured. 

In setup 2 (Fig.\,\ref{Fig:setup2}) the two SiPMs in D1 and D2 are readout independently. 
The corresponding signals C1$\ldots$C4 are processed in a similar way as before; 
four CFD signals F1$\ldots$F4 are generated in DSO. The trigger signals are tr\ =\ (C1)C2C3C4\ =\ (D1)D2, i.e. coincidence of all four signals, each with applied detection threshold, taken on the leading edge of the signal C1, and tr\ =\ (C1)C2\ =\ D1 for calibration and measurements, respectively. The sampling rate of the DSO is 20\,GS/s. The detection times are calculated as
$t_1 = (t_{\rm F1} + t_{\rm F2})/2$ and $t_2 = (t_{\rm F3} + t_{\rm F4})/2$. Such averaging allows compensating for the particle hit position and improving the time resolution of D1 and D2 compared to the setup 1. The use of setup 1 is justified, on the other hand, by a possibility to have the DSO sampling rate of 40\,GS/s and by a more simple and faster re-tuning of the whole signal processing chain, when the detector D2 is changed.

\subsection{Voltage Amplifier}

\begin{figure}[!htb]
\centering
\includegraphics[width=1.0\columnwidth,clip]{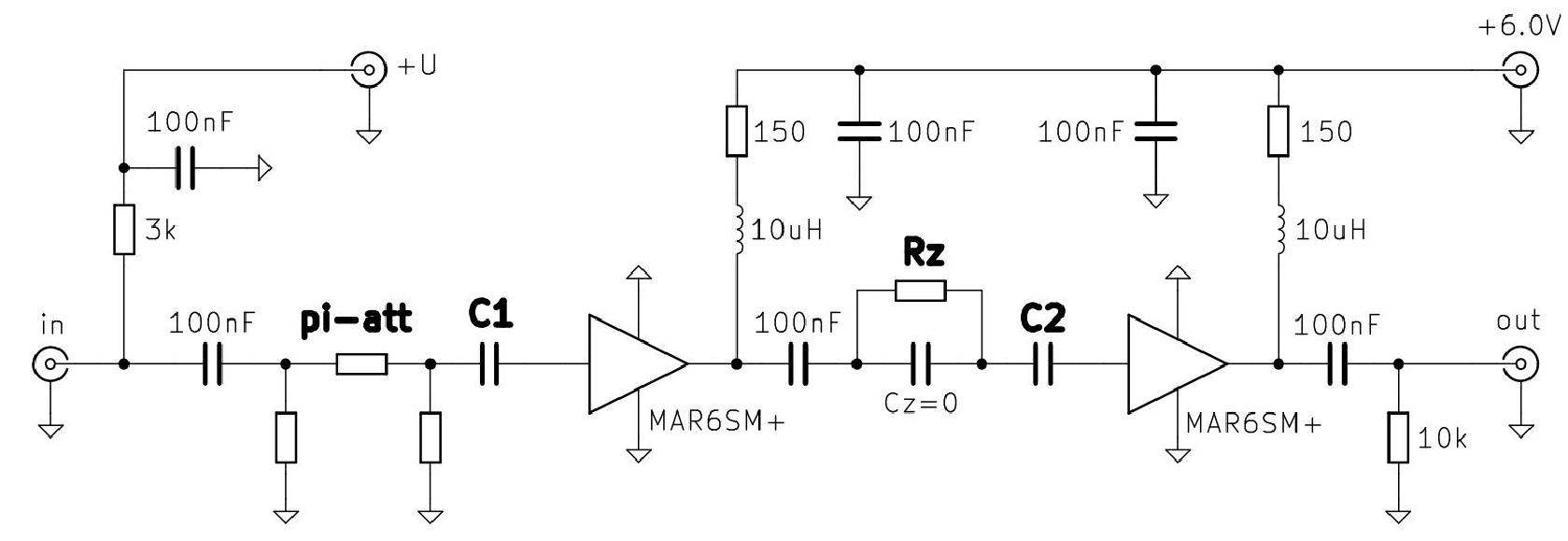}
\caption{
Biasing and the amplifier circuits. The tunable parameters to adjust the gain and the signal pulse-shape are: pi-att, C1, Rz, C2.
}
\label{Fig:amp-scheme}
\end{figure}

Here, as also in other applications \cite{Stoykov12a,Cattaneo14,Amato17,Rostomyan21}, we use a common approach for taking the signals from detectors. In this approach a single SiPM or their group is connected to a bias power supply and to an amplifier via a single 50\,Ohm cable. A typical scheme of the biasing and the amplifier circuits is shown in Fig.\,\ref{Fig:amp-scheme}. A two-stage non-inverting voltage amplifier is built around MAR6SM+ monolithic amplifier chips from Mini-Circuits. It has 50\,Ohm input impedance, adjustable gain and signal pulse shape. 

\subsection{Constant Fraction Discriminator}

\begin{figure}[!htb]
\centering
\includegraphics[width=1.0\columnwidth,clip]{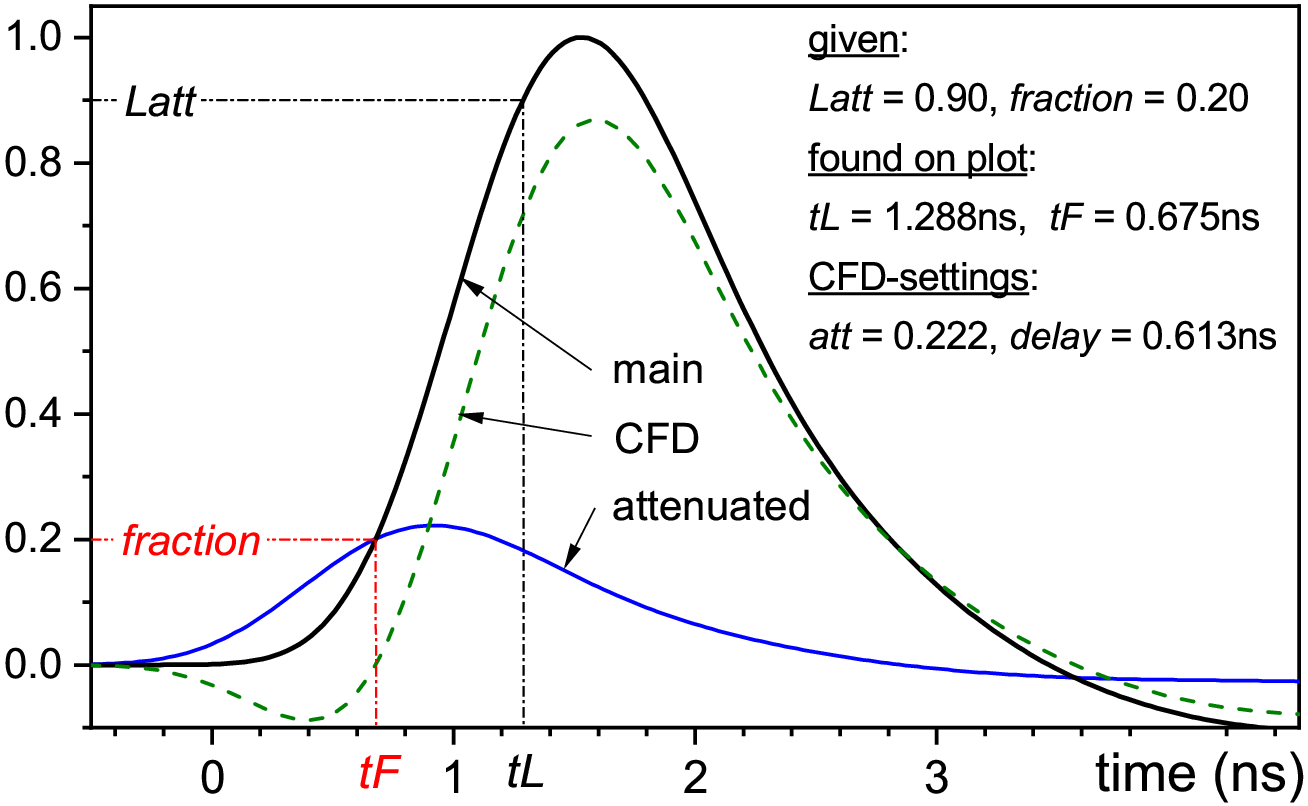}
\caption{
Explanation of the CFD algorithm. The main signal is positive and starts at time-zero. The attenuated signal is the main multiplied by $att$ and shifted backward by $delay$. The CFD signal is calculated as main minus attenuated. The zero-line crossing of the CFD signal gives the trigger time $tF$. The levels of the main and attenuated signals at this time are $fraction$ and $Latt$, respectively. To ensure triggering at the given levels, the CFD settings $att$ and $delay$ are calculated as: $att = fraction/Latt$,  $delay = tL – tF$. Note, that even that the level $Latt$ belongs to the attenuated signal, for the purpose of finding the time $tL$ it is taken on the main signal.
The main signal, used here as an example, is a reverted averaged waveform obtained with a BC422 black scintillator cube read out by two Broadcom SiPMs connected in series. 
The amplifier settings are: 
pi-att\ =\ 6\,dB, C1\ =\ 5.6\,pF, Rz\ =\ 470\,Ohm, C2\ =\ 22\,pF.  
The signal rise time is 0.75\,ns.
}
\label{Fig:CFD}
\end{figure}

As the pulse-height distributions of D1 and D2 signals are rather broad, we use a constant fraction discriminator (CFD) technique \cite{Leo} to correct for the signal walk.  In a CFD the trigger time corresponds to a given fraction of the signal and is independent of its amplitude as long as the signal shape is preserved. The constant fraction algorithm is realized in the DSO software and is explained in Fig.\,\ref{Fig:CFD}. The input signal (we take it to be positive and starting at time zero) is split in two signals: main and attenuated. The main signal is the unchanged input one. The attenuated signal is the main multiplied by a factor $att$ and shifted backward by introducing a negative time offset, referred to as $delay$. The crossing of the main and attenuated signals occurs at their levels $fraction$ and $Latt$, respectively. 
The corresponding time $tF$ is the CFD trigger time. In practice $tF$ is found by zero-crossing of a CFD signal, calculated as main minus attenuated. The input parameters used for tuning the CFD are $fraction$ and $Latt$, i.e. the levels of main and attenuated signals, at which the trigger will occur. In the classical description of the CFD algorithm \cite{Leo} it is required the crossing of the main and attenuated signals to occur at the maximum of the attenuated signal (the case of $Latt = 1$). In this work we set $Latt = 0.9$.
By using the averaged signal waveform we find the trigger time $tF$, corresponding to the level $fraction$, and the time $tL$, corresponding to the level $Latt$ taken on the main signal. Note, that even that the level $Latt$ belongs to the attenuated signal, for the purpose of finding the time $tL$ it is taken on the main signal. The CFD settings $att$ and $delay$ to ensure triggering at the given levels are: $att = fraction/Latt$, $delay = tL - tF$. 

\section{Measurements}

The following measurements are aimed at obtaining the time resolution of the detector under test D2 at varying experimental conditions, including its construction (scintillators, SiPMs) and the signal processing. This is achieved by measuring the coincidence time resolution between D2 and the reference detector D1 and subtracting the time resolution of D1, found in calibration measurements.

\subsection{Calibration}

\begin{figure}[!htb]
\centering
\includegraphics[width=1.0\columnwidth,clip]{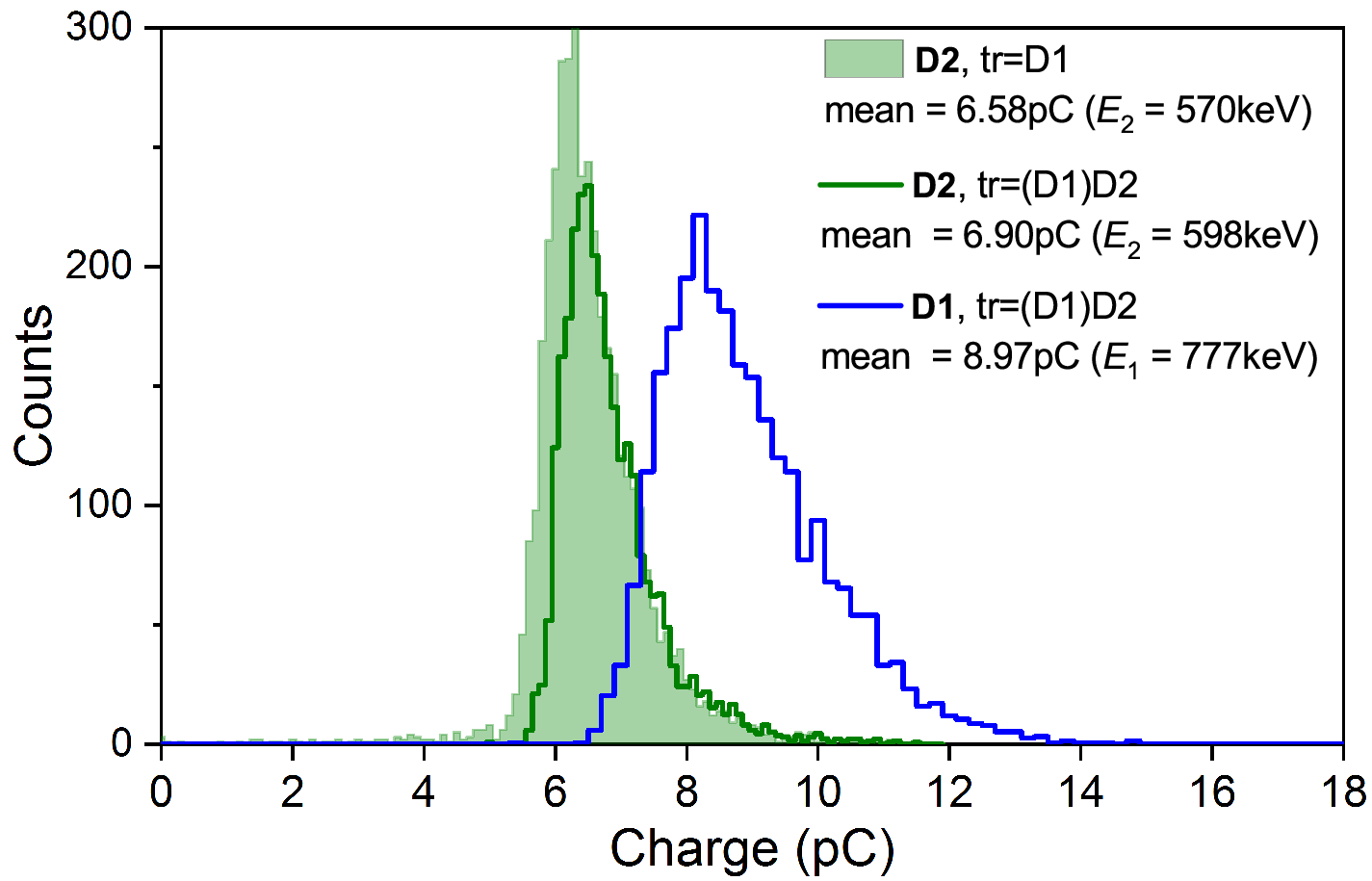}
\caption{
Energy calibration in setup 2. The detectors D1 and D2 are identical: BC422 cubes wrapped in Teflon with Broadcom SiPMs coupled via optical cement. Shown are charge spectra of D1 and D2 signals at chosen detection thresholds with tr\,=\,(D1)D2.
The full spectrum of D2 taken with tr\,=\,D1 is also shown. The mean signal charge of 6.58\,pC in this spectrum corresponds to the mean deposited energy $E_2 = 570$\,keV (see text). The established correspondence between the measured charge and the energy allows finding the mean deposited energies $E_1$ and $E_2$ at tr\,=\,(D1)D2.
}
\label{Fig:cal-energy}
\end{figure}

For calibration measurements the detector under test D2 is chosen to be identical with the reference detector D1: BC422 cube wrapped in Teflon with Broadcom SiPMs coupled via optical cement. 
Figure~\ref{Fig:cal-energy} shows charge spectra of D1 and D2 signals taken with setup 2 under the trigger condition tr\,=\,(D1)D2, i.e. by setting a certain detection threshold for each channel. The charge is obtained by integrating only the negative part of the signal waveform (positive part for the inverted signal shown in Fig.\,\ref{Fig:CFD}).
The threshold value for D2 is chosen just high enough to cut the low-energy tail of the spectrum. The threshold for D1 is increased aiming at the relative width of D1 spectrum (ratio of the standard deviation to the mean) to approach that of D2. The threshold increase for D1 is, however, constrained by the requirement of having the coincidence count rate with the given radioactive source above 10\,cps. The full spectrum of D2 signals taken with tr\,=\,D1 is also measured and shown in Fig.\,\ref{Fig:cal-energy}. 
The mean charge of 6.58\,pC in this spectrum corresponds to the mean deposited energy of 570\,keV 
in the scintillator. This energy estimate comes from simulations \cite{Cattaneo14,Stoykov12b}, where for similar detector setups with 2\,mm and 5\,mm thick plastic scintillators the mean deposited energies were obtained as 380\,keV and 950\,keV, respectively. 
The established correspondence between the measured charge and the energy allows finding the mean deposited energies $E_1$ and $E_2$ at tr\,=\,(D1)D2.

\begin{figure}[!htb]
\centering
\includegraphics[width=1.0\columnwidth,clip]{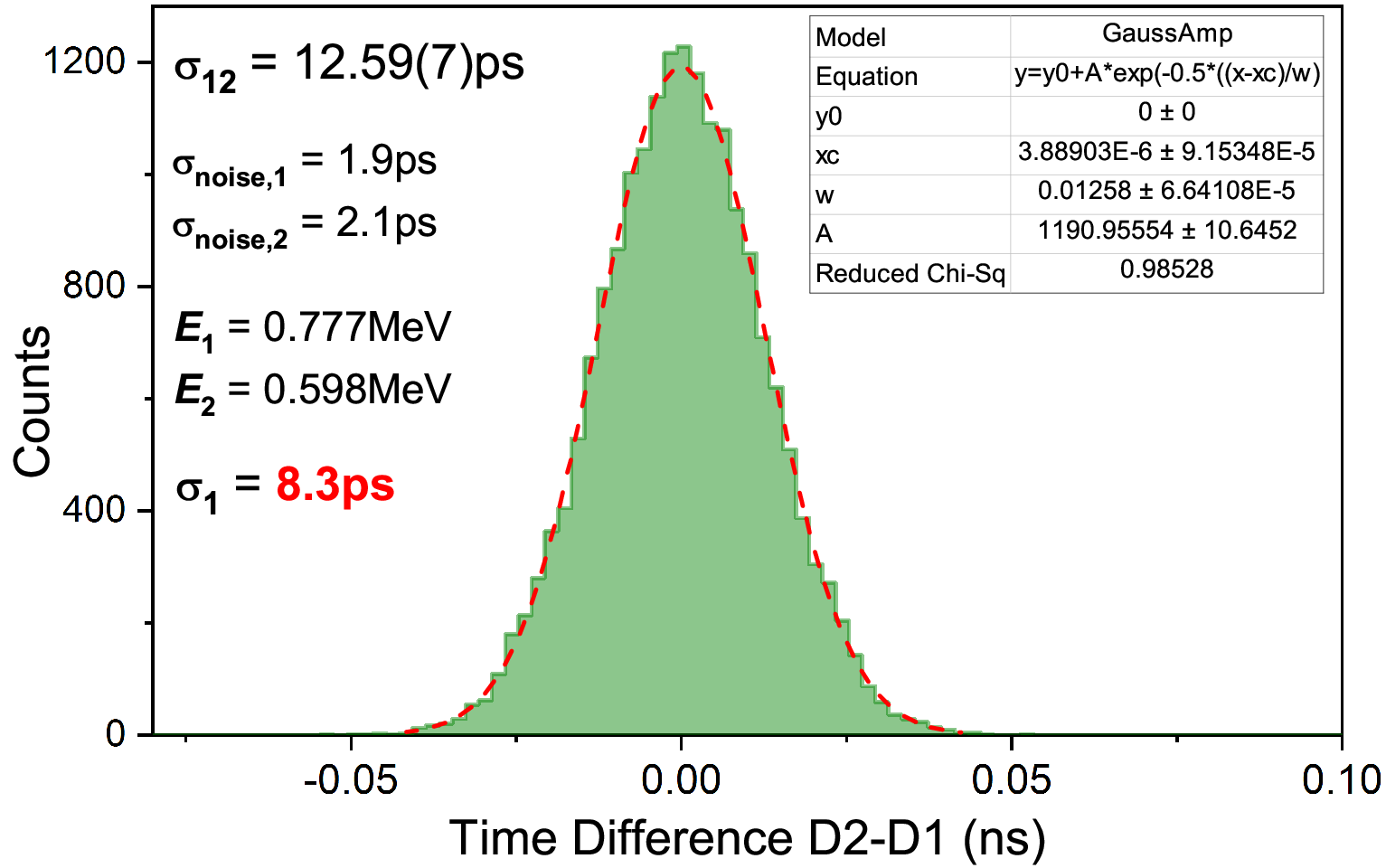}
\caption{
Time calibration in setup 2. Time-difference spectra between D2 and D1 are taken with tr\,=\,(D1)D2 and the detection thresholds same as for Fig.\,\ref{Fig:cal-energy}. The Gaussian fit to the data (dashed line) gives the coincidence time resolution $\sigma_{12} = 12.6$\,ps. Calculated from this value using (\ref{Eq:cal-sigma1}), the time resolution of the reference detector is $\sigma_1 = 8.3$\,ps.
}
\label{Fig:cal-time}
\end{figure}

Figure~\ref{Fig:cal-time} shows the time-difference spectra between D2 and D1 taken with tr\,=\,(D1)D2 and the detection thresholds same as for Fig.\,\ref{Fig:cal-energy}. By fitting the Gaussian function to this distribution we get the coincidence time resolution between the two detectors $\sigma_{12}$. From this value and the mean deposited energies $E_1$ and $E_2$ the time resolution $\sigma_1$ of the reference detector is calculated as:

\begin{equation}
\sigma_1 = \sqrt{(\sigma_{12}^2 - \sigma_{\rm noise,1}^2 - \sigma_{\rm noise,2}^2)
\,\frac{E_2}{E_1 + E_2} + \sigma_{\rm noise,1}^2}\ ,
\label{Eq:cal-sigma1}
\end{equation}
where $\sigma_{\rm noise,1}$  and $\sigma_{\rm noise,2}$ are the electronic noise contributions to the time resolution of D1 and D2, obtained from the measured time jitter $\sigma_{\rm noise,F}$  of the CFD signals at their zero-crossing. For setup 1: $\sigma_{\rm noise,1} = \sigma_{\rm noise,F1}$ and
$\sigma_{\rm noise,2} = \sigma_{\rm noise,F2}$.
For setup 2: $\sigma_{\rm noise,1} = \frac{1}{\sqrt{2}} \sigma_{\rm noise,F1}$  and
$\sigma_{\rm noise,2} = \frac{1}{\sqrt{2}} \sigma_{\rm noise,F3}$. In case of setup 2 the noise in channel F2 is equal to that in F1, and in channel F4 equal to F3: accordingly only the time jitters $\sigma_{\rm noise,F1}$  and
$\sigma_{\rm noise,F3}$  are measured in this setup.
The values $\sigma_{12}$  (CTR) and $\sigma_1$  (time resolution of the reference detector), obtained in calibration measurements, are listed in Table~\ref{Table:calibration}.

\begin{table}[!htb]
\centering
\renewcommand{\arraystretch}{1.1}
\caption{
Coincidence time resolution $\sigma_{12}$ and the time resolution of the reference detector $\sigma_1$ obtained in calibration measurements with setup 1 and setup 2
}
\begin{tabular}{ccc}
\hline
\lower 5pt \hbox{}
  & \lower 5pt \hbox{$\sigma_{12}$,\,ps}
  & \lower 5pt \hbox{$\sigma_1$,\,ps} \\[10pt]
\hline
setup 1	& 14.3 & 9.8 \\
setup 2	& 12.6 & 8.3 \\
\hline
\end{tabular}
\label{Table:calibration}
\end{table}

\subsection{BC422 cubes}

In the following we measure the time resolution of the detector D2 under different conditions (SiPM, signal processing) and calculate the time resolution per 1\,MeV detected energy as:

\begin{equation}
\sigma = \sqrt{\sigma_{12}^2 - \sigma_1^2} \ ,
\label{Eq:meas-sigma2}
\end{equation}

\begin{equation}
\sigma_{\rm 1MeV} = \sqrt{k E (\sigma^2 - \sigma_{\rm noise}^2)} \ ,
\label{Eq:meas-sigma1MeV}
\end{equation}
where $k$, $E$, and $\sigma$ are the light collection, the mean deposited energy, and the measured time resolution of D2, respectively; $\sigma_{\rm noise}$ is the electronic noise contribution to $\sigma$. 
In contrast to calibration, no detection threshold is applied to the signals of D2 during the measurements: 
the mean deposited energy in D2 is $E = 570$\,keV. 
Note, that here and further on we omit the sub-indexes 2 under the parameters referring solely to D2. Also, as all the changes from now on are done only to the detector D2, all references to the detector characteristics and to the parameters of the signal processing imply only this detector. 

\begin{figure}[!htb]
\centering
\includegraphics[width=1.0\columnwidth,clip]{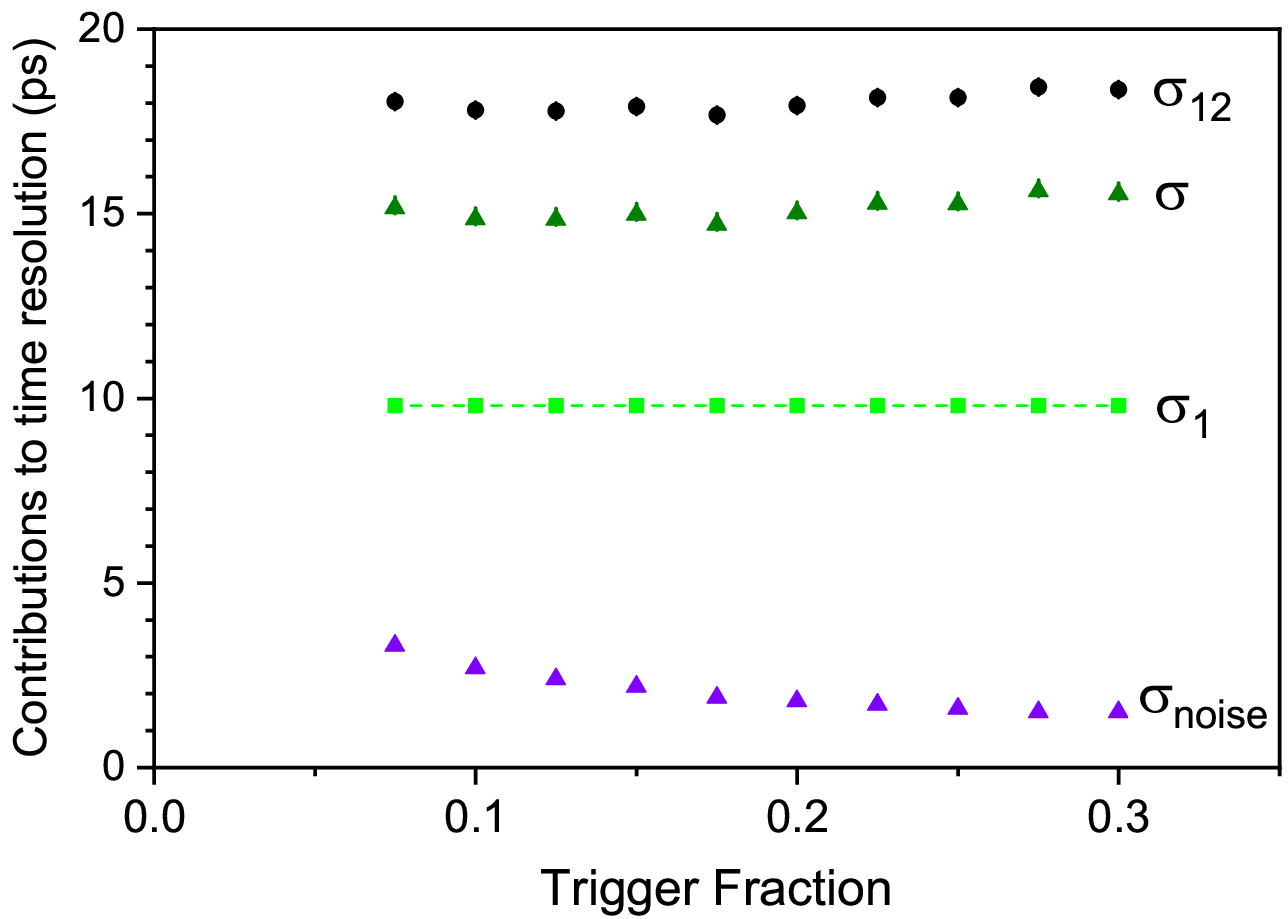}
\caption{
Contributions to the measured time resolution as a function of the trigger fraction. Sub-indexes under the parameters referring to D2 are omitted. Experimental conditions: setup 1, Broadcom, BC422 (cube, black, cement), signal rise time 0.75\,ns, bandwidth 580\,MHz, sampling rate 40\,GS/s.
}
\label{Fig:contributions}
\end{figure}

\begin{figure}[!htb]
\centering
\includegraphics[width=1.0\columnwidth,clip]{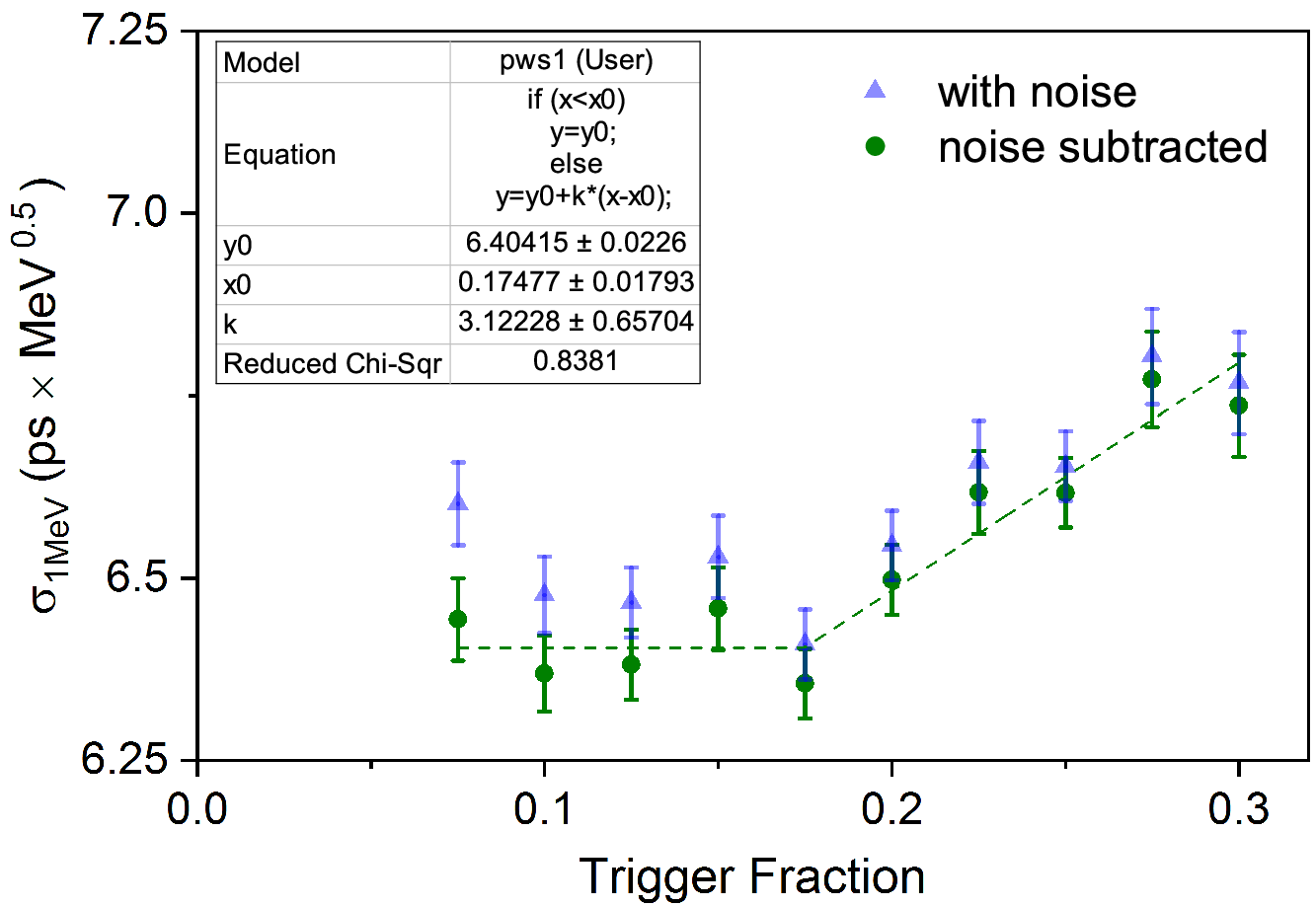}
\caption{
Dependencies of $\sigma_{\rm 1MeV}$ on the trigger fraction extracted using (\ref{Eq:meas-sigma1MeV}) from the data in Fig.\,\ref{Fig:contributions}. The dashed line is a fit to the data using a combination of a constant line and a linear function with the fit results given in the plot.
}
\label{Fig:noise-subtraction}
\end{figure}

Figure~\ref{Fig:contributions} shows contributions to the measured time resolution as a function of the trigger fraction obtained with setup\ 1 and Broadcom SiPMs. 
The corresponding dependencies of $\sigma_{\rm 1MeV}$ with and without noise subtraction are shown in Fig.\,\ref{Fig:noise-subtraction}. The noise contribution to $\sigma_{\rm 1MeV}$ increases with decreasing the trigger fraction thus limiting its minimum accessible value to about 0.07. Further on all the data on $\sigma_{\rm 1MeV}$, unless explicitly mentioned, are given with the noise contribution subtracted.

The light collection $k$ in a painted black ($3 \times 3 \times 3$)\,mm$^3$ cube scintillator is taken as 1/3, i.e. it is assumed that the light source is distributed uniformly over the scintillator volume and all six faces of the scintillator, four painted black and two with coupled SiPMs, are fully absorbing. 
Accordingly, the detected energy $E_{\rm det} = k E$ in such detectors is:
$E_{\rm det} = 1⁄3 \times 570\,{\rm keV} \approx 190\,{\rm keV}$. 
The light collection in a Teflon wrapped scintillator with two SiPMs coupled via optical grease, measured relative to the black one, is found to be 0.88. This measurement is done with Broadcom SiPMs at a low overvoltage to avoid a systematic shift due to the optical crosstalk. Even though we do not use this value in our analysis, it serves as a cross-check for the validity of the above estimate for the black cubes. As the light collection can`t exceed 1, the maximum possible negative systematic shift in estimating the detected energy is 12\,\% and, accordingly, 6\,\% in $\sigma_{\rm 1MeV}$. 
Similar measurement with a SiPM attached only to one scintillator face and other five being wrapped in Teflon, yields the result of $k = 0.74$ in agreement with \cite{Gundacker20}.

\begin{figure}[!htb]
\centering
\includegraphics[width=0.95\columnwidth,clip]{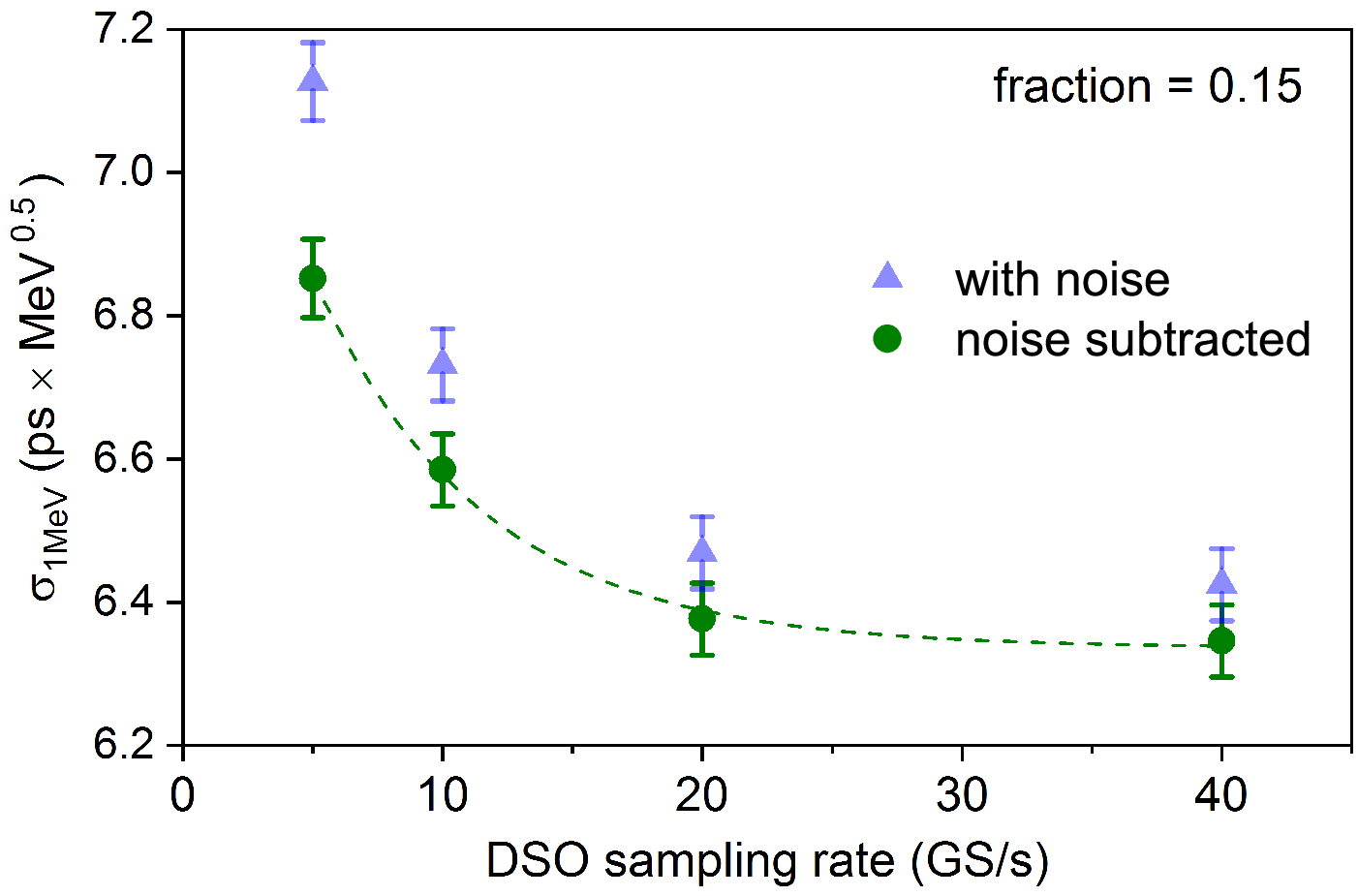}
\caption{
Time resolution $\sigma_{\rm 1MeV}$ as a function of DSO sampling rate. The dashed line is drawn to guide an eye. Experimental conditions: setup 1, Broadcom, BC422 (cube, black, cement), rise time 0.75\,ns,
bandwidth $\approx 600$\,MHz.
}
\label{Fig:s-rate}
\end{figure}

As in practical applications one often needs to take compromises between the detector performance on the one hand, and its technical complexity and the cost on the other, the sampling rate as high as 40\,GS/s, used in this work, might not be possible. Figure~\ref{Fig:s-rate} shows degradation of $\sigma_{\rm 1MeV}$ at lower sampling rates. This measurement, as most of the others in this work, is done with Broadcom SiPMs, which provide the best timing performance. Down to 5\,GS/s the value of $\sigma_{\rm 1MeV}$ increases by about 8\,\%. Also notice an increased noise contribution with decreasing the sampling rate. This comes from the fact that at decreasing the sampling rate we, to keep the bandwidth constant, also decrease the strength (number of bits) of the ERes filter, which makes the noise suppression less efficient.

\begin{figure}[!htb]
\centering
\includegraphics[width=0.95\columnwidth,clip]{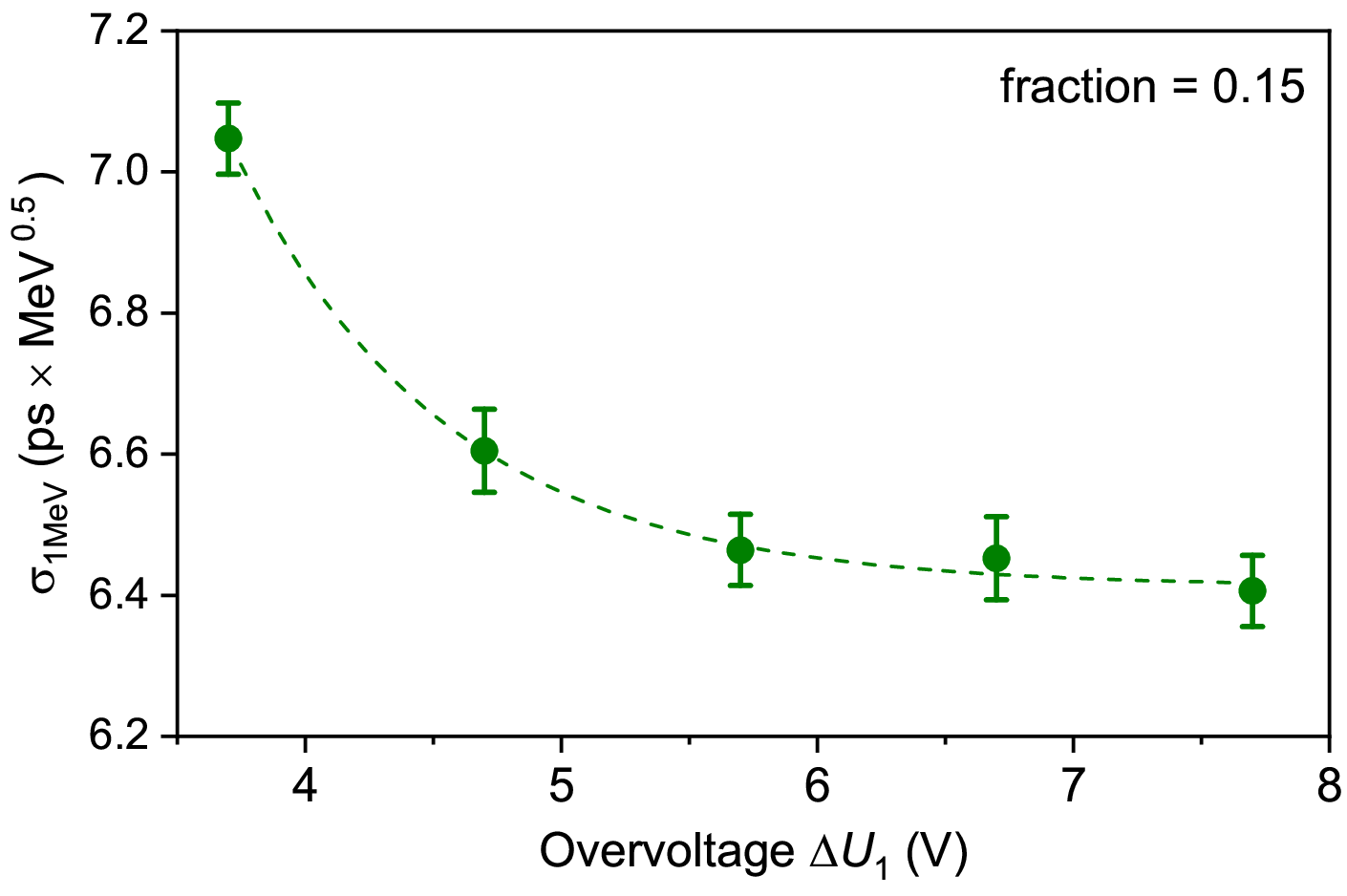}
\caption{
Time resolution $\sigma_{\rm 1MeV}$ as a function of overvoltage measured with a detector using Broadcom SiPMs. The dashed line is drawn to guide an eye. The chosen operation point is at $\Delta U_1 = 6.7$\,V. Experimental conditions: setup 1, Broadcom, BC422 (cube, black, cement), signal rise time 0.75\,ns, bandwidth 580\,MHz, sampling rate 40\,GS/s.
}
\label{Fig:dU}
\end{figure}

An overvoltage dependence of $\sigma_{\rm 1MeV}$ is measured only with Broadcom SiPMs 
(see Fig.\,\ref{Fig:dU}).
It shows initially some decrease with increasing $\Delta U_1$ and saturates at larger $\Delta U_1$ values (both PDE and SPTR saturate \cite{Gundacker20}). With other SiPMs we rely on the chosen overvoltages being sufficiently high to ensure a close-to-optimum performance.

\begin{figure}[!htb]
\centering
\includegraphics[width=1.0\columnwidth,clip]{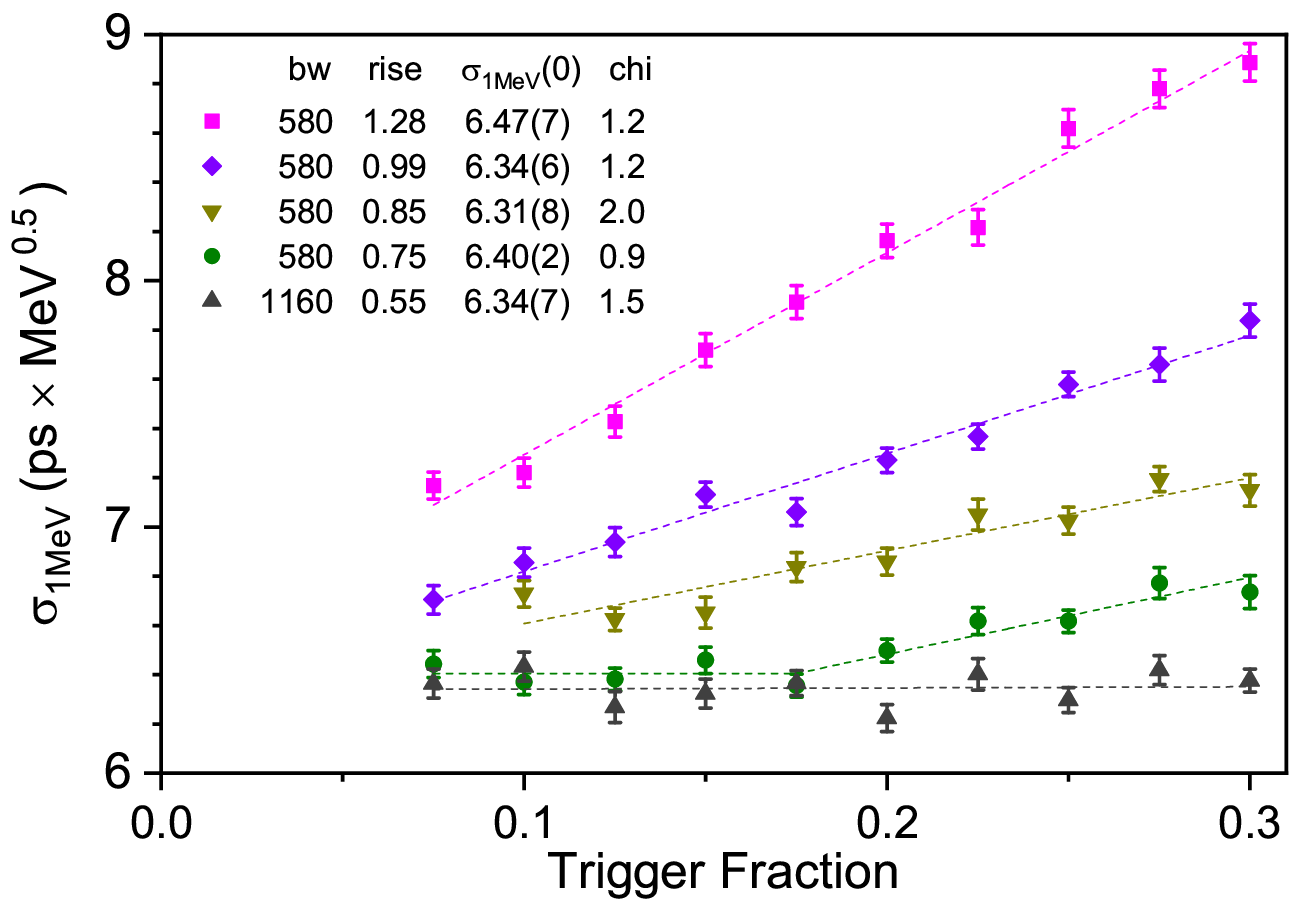}
\caption{
Time resolution $\sigma_{\rm 1MeV}$  as a function of trigger fraction at different settings of the amplifier, resulting in different signal rise times. The dashed lines are fits to the data using either a linear function or a combination of a linear function applied above certain fraction value and a constant line below this value.
The DSO bandwidth (MHz), the signal rise time (ns),
the minimum time resolution $\sigma_{\rm 1MeV}(0)$ (ps$\times$MeV$^{0.5}$)
and, characterizing the quality of the fit, reduced chi-square value are indicated on the plot. The measurement with DSO bandwidth 1160\,MHz is done only once in this work. Experimental conditions: setup 1, Broadcom, BC422 (cube, black, cement), sampling rate 40\,GS/s.
}
\label{Fig:sigma1MeV=f(amp)}
\end{figure}

Figure~\ref{Fig:sigma1MeV=f(amp)} shows dependencies of $\sigma_{\rm 1MeV}$ on the trigger fraction at different settings of the amplifier providing different signal rise times. 
For a given fraction, the value of $\sigma_{\rm 1MeV}$ is smaller the shorter is the signal rise time. 
Extrapolated to zero fraction $\sigma_{\rm 1MeV}(0)$ seems to be independent on the rise time.

\begin{figure}[!htb]
\centering
\includegraphics[width=1.0\columnwidth,clip]{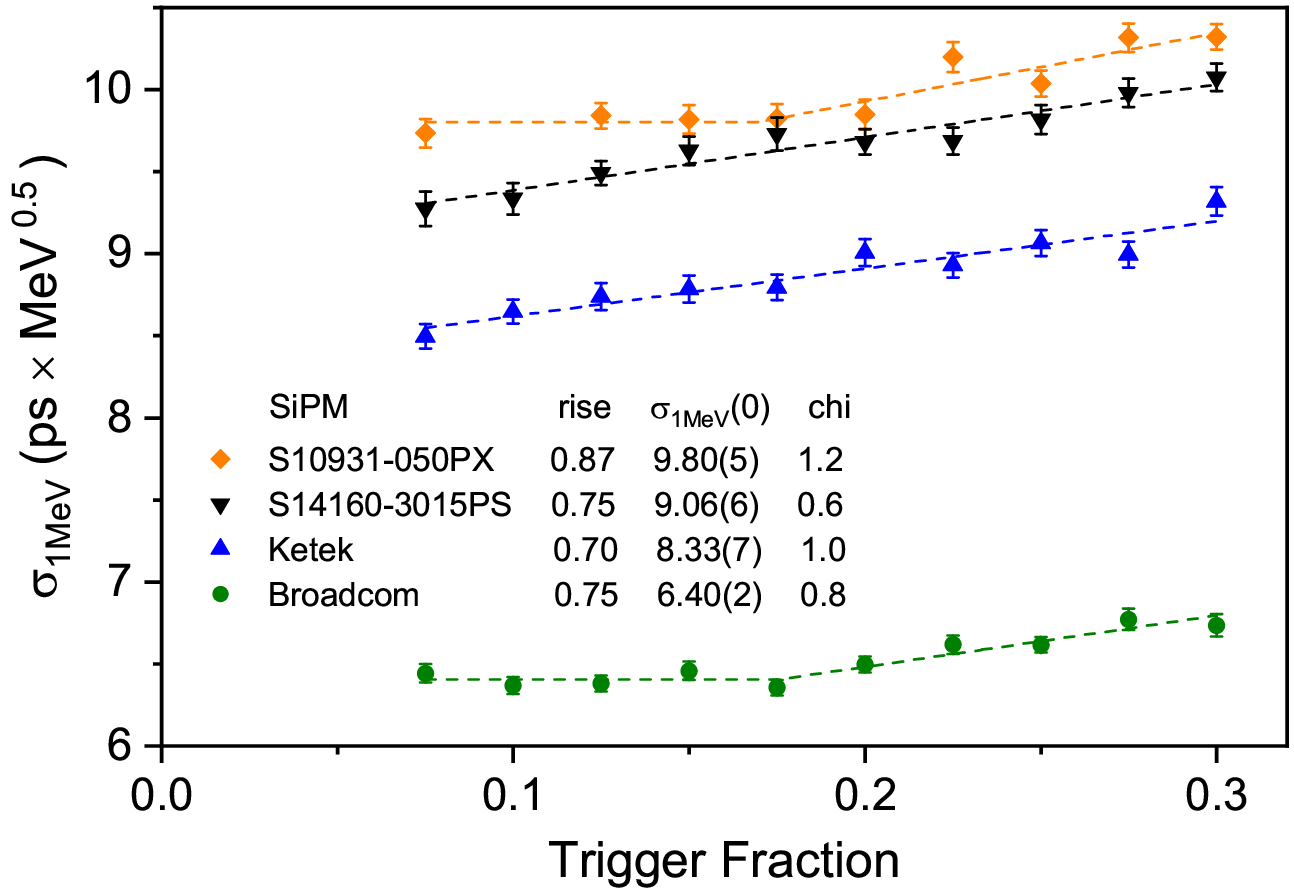}
\caption{
Time resolution $\sigma_{\rm 1MeV}$ as a function of trigger fraction obtained with different SiPMs. For clarity, not all tested SiPMs are shown. The dashed lines are fits to the data explained in 
Fig.\,\ref{Fig:sigma1MeV=f(amp)}. The SiPM type, signal rise time (ns), $\sigma_{\rm 1MeV}(0)$ (ps$\times$MeV$^{0.5}$), and reduced chi-square value, are indicated on the plot. Experimental conditions: setup 1, BC422 (cube, black, cement), bandwidth 580\,MHz,
sampling rate 40\,GS/s.
}
\label{Fig:sigma1MeV=f(SiPM)}
\end{figure}

Figure~\ref{Fig:sigma1MeV=f(SiPM)} shows dependencies of $\sigma_{\rm 1MeV}$  on the trigger fraction obtained with different SiPMs. Table~\ref{Table:summary-cube} summarizes the data on $\sigma_{\rm 1MeV}(0)$ for all tested SiPMs. Among SiPMs available for this study, the one from Broadcom is by far the best for timing applications using BC422 scintillator with
$\sigma_{\rm 1MeV}(0) = 6.4$\,ps$\times$MeV$^{0.5}$. The oldest Hamamatsu SiPM S10931-050PX shows the worst time resolution saturating at the level of 9.8\,ps$\times$MeV$^{0.5}$. This result, however, is substantially better compared to that obtained in \cite{Stoykov12b} with the same type device. As a consistency check we repeated the measurement with this SiPM under the conditions of \cite{Stoykov12b} (signal rise time 1.3\,ns,
$\Delta U_1 = 1.0$\,V, $Latt = 1.0$, $fraction = 0.2$) and obtained $\sigma_{\rm 1MeV}$ values in the range ($14.8 - 15.8$)\,ps$\times$MeV$^{0.5}$, consistent with \cite{Stoykov12b}.

\begin{table}[!htb]
\centering
\renewcommand{\arraystretch}{1.1}
\caption{
Summary on the minimum time resolution $\sigma_{\rm 1MeV}(0)$ for all SiPMs available for the current study measured with setup 1
}
\begin{tabular}{cc}
\hline
\lower 5pt \hbox{SiPM} & \lower 5pt \hbox{$\sigma_{\rm 1MeV}(0)$,\,ps$\times$MeV$^{0.5}$} \\[10pt]
\hline
\phantom{0}Ham S10931-050PX	& $9.80 \pm 0.05$ \\
\phantom{00}Ham S12572-025P		& $9.46 \pm	0.05$ \\
\phantom{}Ham S14160-3015PS		& $9.06 \pm 0.06$ \\
\phantom{0000000009}Advansid	& $8.84 \pm	0.04$ \\
\phantom{0000000000000}Ketek	& $8.33 \pm	0.07$ \\
\phantom{}Ham S13360-3050PE		& $8.16 \pm	0.04$ \\
\phantom{}Ham S14160-3050HS	& $7.83 \pm	0.07$ \\
\phantom{0000000000}Broadcom	& $6.40 \pm	0.02$ \\
\hline
\end{tabular}
\label{Table:summary-cube}
\end{table}

\begin{figure}[!htb]
\centering
\includegraphics[width=1.0\columnwidth,clip]{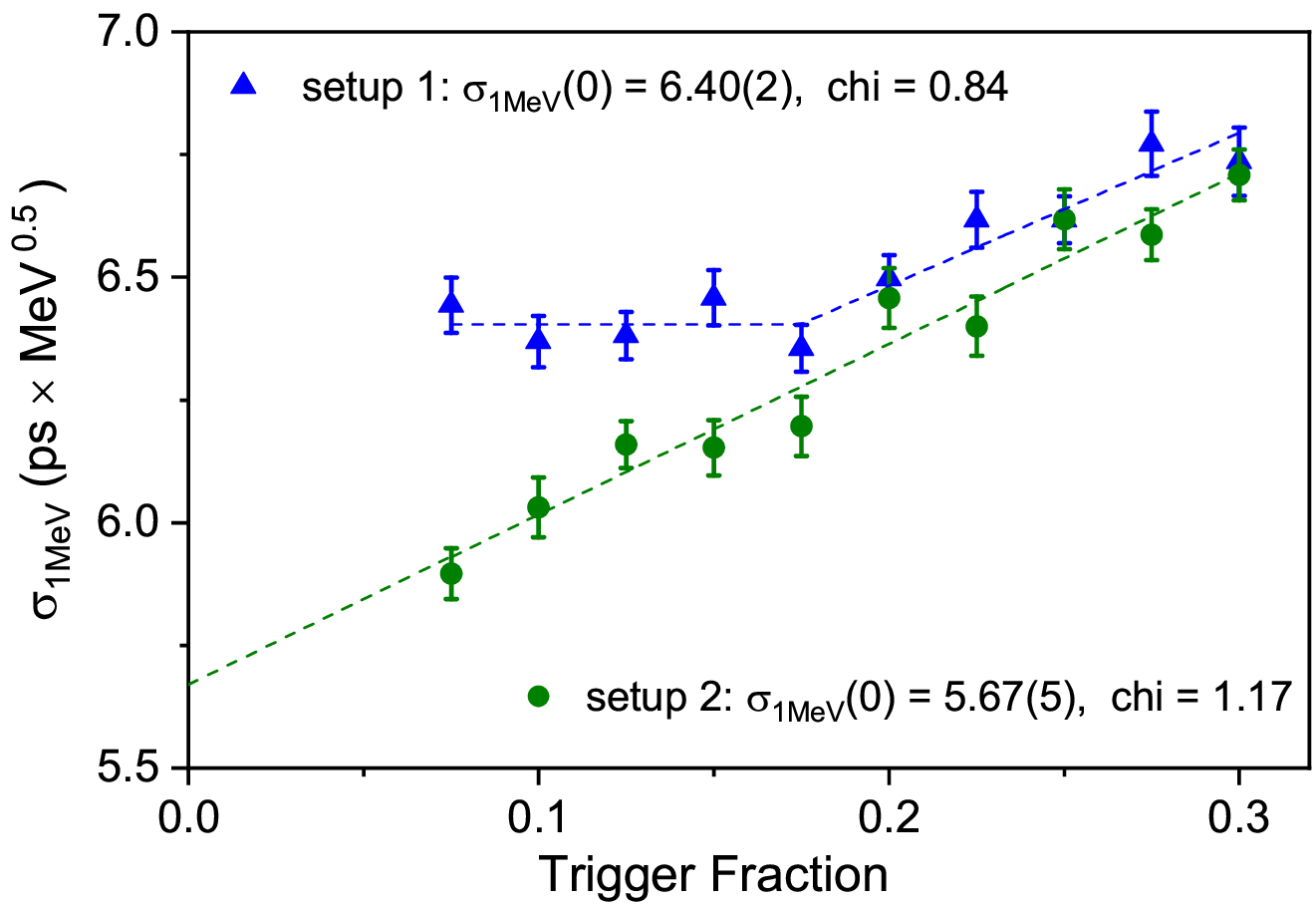}
\caption{
Time resolution $\sigma_{\rm 1MeV}$ as a function of trigger fraction obtained with Broadcom SiPMs using setup 1 and setup 2. The dashed lines are fits to the data explained in Fig.\,\ref{Fig:sigma1MeV=f(amp)}.
Obtained in the fit $\sigma_{\rm 1MeV}(0)$ in ps$\times$MeV$^{0.5}$ and the reduced chi-square value are indicated on the plot. Experimental conditions setup 1: Broadcom, BC422 (cube, black, cement), rise time 0.75\,ns, bandwidth 580\,MHz, sampling rate 40\,GS/s. Experimental conditions setup 2: Broadcom, BC422 (cube, black, grease), rise time 0.77\,ns, bandwidth 580\,MHz, sampling rate 20\,GS/s.
}
\label{Fig:sigma1MeV=f(setup)}
\end{figure}

As in the setup 1, used up to now, there is no compensation for the particle hit position, some potential for improving the time resolution is still present. To check this we repeat the measurements with Broadcom SiPMs using setup 2 and keeping all the relevant settings the same as before.
No saturation of $\sigma_{\rm 1MeV}$ and its improvement by another 10\,\% to
$\sigma_{\rm 1MeV}(0) = 5.7$\,ps$\times$MeV$^{0.5}$ is observed 
(see Fig.\,\ref{Fig:sigma1MeV=f(setup)}). This result approaches that of \cite{Gundacker20} obtained, however, with a SiPM with presumably better timing performance.

\subsection{BC422 stripe}
To the cumulative time spread $\sigma_{\rm ts}$ in (\ref{Eq:sigma1MeVmin}) contribute both the light collection and the SiPM response formation processes. 
This means that in practice one might face a situation when all advantages of using a fast scintillator read out by a low SPTR value SiPM are gone at changing from a tiny test detector to a real size one, having substantially larger dimensions and, accordingly, a larger PTS. To clarify this question, we perform measurements with a BC422 scintillator stripe of dimensions ($100 \times 14 \times 4$)\,mm$^3$ read out by 3 Broadcom SiPMs attached to each ($14 \times 4$)\,mm$^2$ face (see Fig. 1).
The SiPMs are connected in series; the coupling to the scintillator is done using optical grease. In terms of maximum dimension of 100\,mm we consider this detector to adequately represent our typical detectors \cite{Stoykov12a,Cattaneo14,Rostomyan21,Amato17}. The measurement is performed with setup 2; the collimated $^{90}$Sr source is positioned at the center of a ($100 \times 14$)\,mm$^2$ face of the scintillator. To obtain the detected energy we subsequently connect to the output only one SiPM out of 3 in each group (the other 2 are short cut) and thus measure the charge spectra. Mean values of the charge, obtained in three such measurements, are summed and compared with the charge measured under the same conditions with BC422 ($3 \times 3 \times 3$)\,mm$^3$ painted black cube, where the mean detected energy equals to 190\,keV. Accordingly, for the stripe we get $E_{\rm det} = 121$\,keV. Taking $E = 760$\,keV for the expected mean deposited energy in 4\,mm thick scintillator \cite{Cattaneo14,Stoykov12b}, for the light collection we get: $k = 0.16$. Often it is convenient, especially for large detectors, to separate different contributions to the light collection as: $k = k_1 k_2$, where $k_1$ is the fraction of the scintillation light transported from the particle interaction point to the scintillator surface, to which the photosensor is attached (light transport), and $k_2$ is the fraction of the transported light actually seen by the photosensor (sensor acceptance). Here we assume a perfect optical contact between the scintillator and the SiPMs, ensuring no reflection on this interface, and take for the sensor acceptance its geometrical value, calculated as the ratio of the total active area of all attached SiPMs to that of the corresponding scintillator surfaces: $k_2 = 0.74$. Accordingly, for the light transport we get: 
$k_1 \approx 0.22$.

\begin{figure}[!htb]
\centering
\includegraphics[width=1.0\columnwidth,clip]{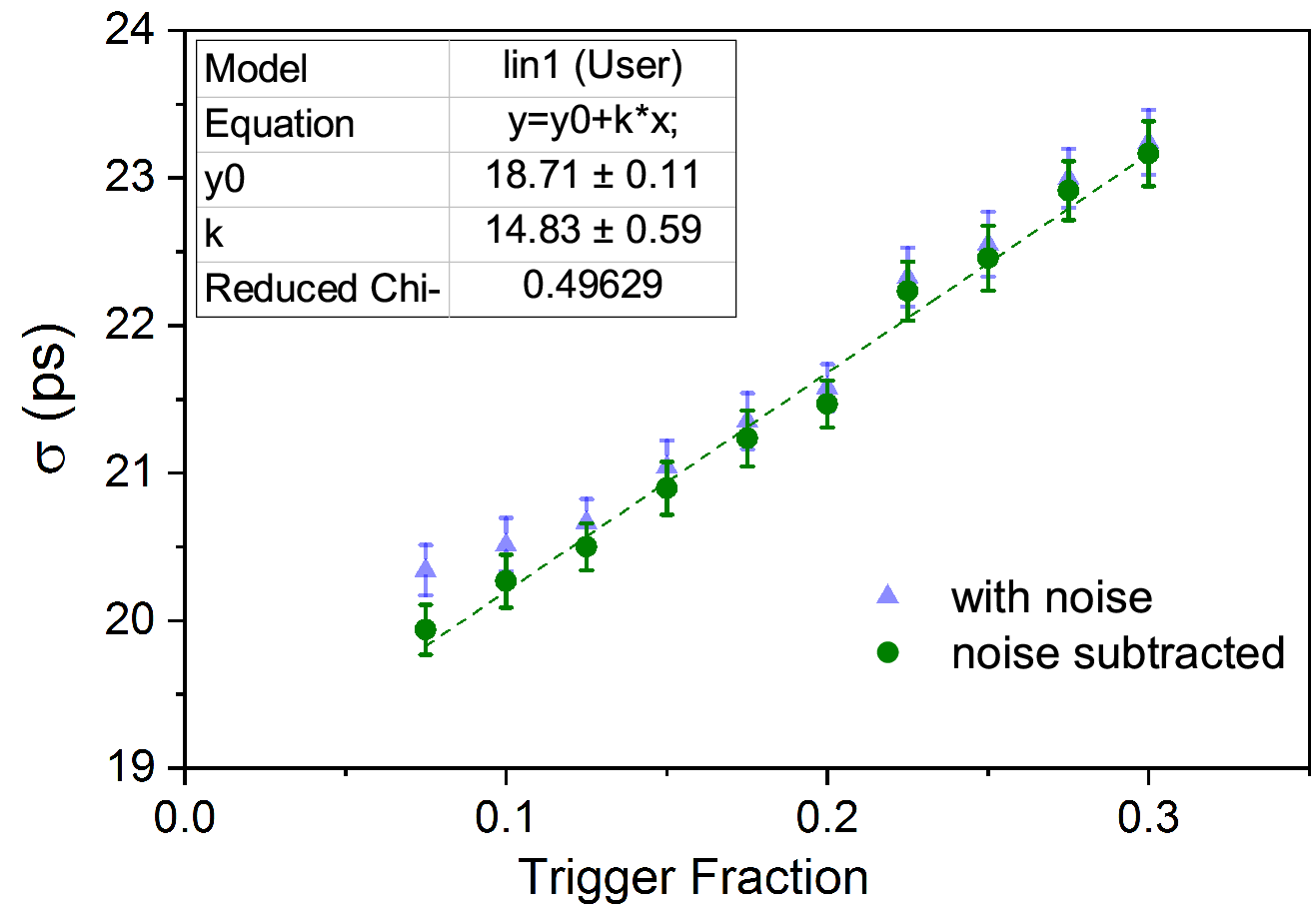}
\caption{
Time resolution $\sigma$ as a function of trigger fraction measured with BC422 stripe depicted 
in Fig.\,\ref{Fig:scint}. The collimated $^{90}$Sr source is positioned at the center of a ($100 \times 14$)\,mm$^2$ face of the scintillator. The dashed line is a linear fit to the data; the results of the fit are given on the plot. Experimental conditions: setup 2, Broadcom (3x, series), BC422 (stripe, grease), rise time 0.83\,ns, bandwidth 580\,MHz,
sampling rate 20\,GS/s.
}
\label{Fig:stripe-sigma}
\end{figure}

\begin{figure}[!htb]
\centering
\includegraphics[width=1.0\columnwidth,clip]{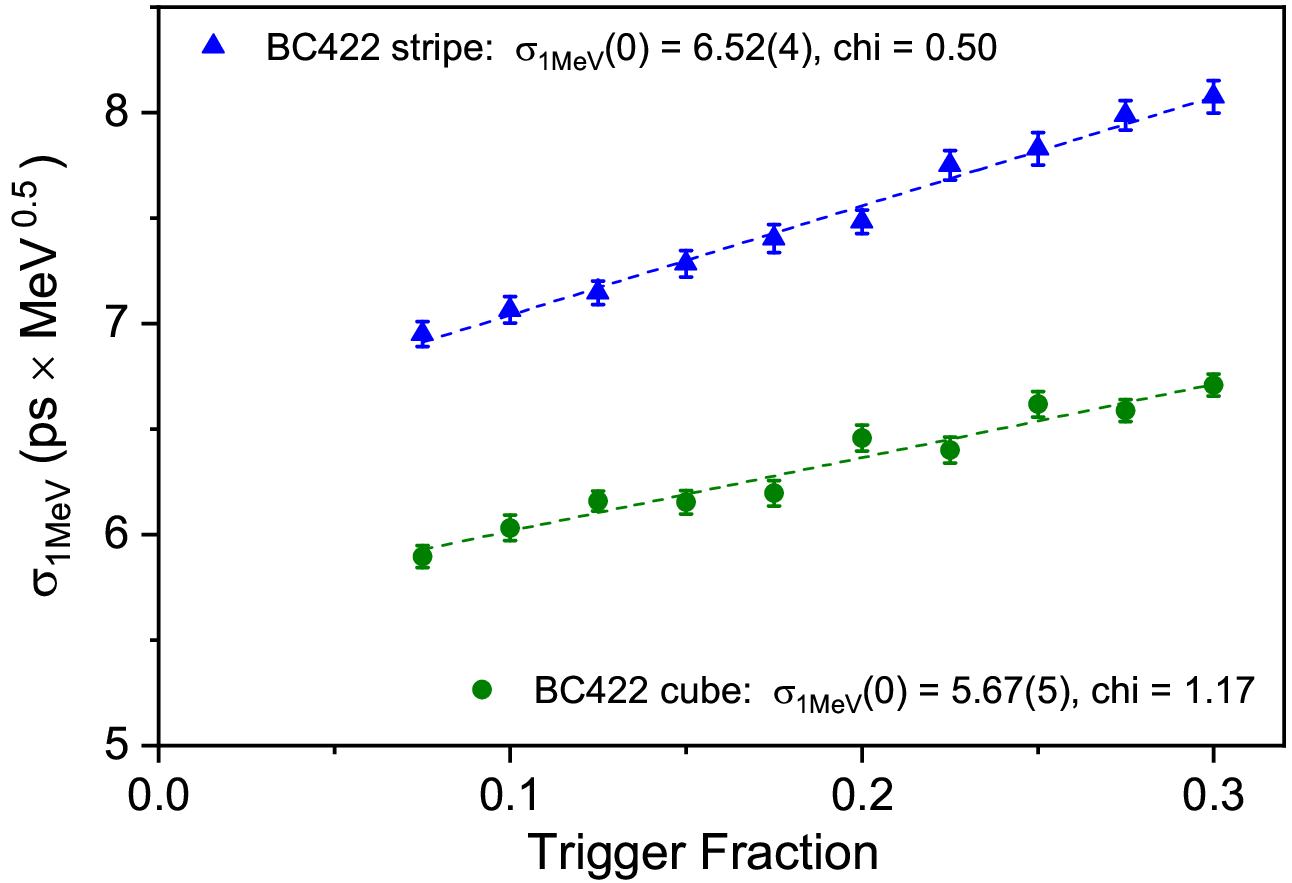}
\caption{
Time resolution $\sigma_{\rm 1MeV}$ as a function of trigger fraction for BC422 stripe compared to that for BC422 black painted cube measured under similar experimental conditions. The dashed lines are linear fits to the data. The $\sigma_{\rm 1MeV}(0)$ (ps$\times$MeV$^{0.5}$) and the reduced chi-square value are indicated on the plot.
Experimental conditions stripe: setup 2, Broadcom (3x, series), BC422 (stripe, grease), rise time 0.83\,ns, bandwidth 580\,MHz, sampling rate 20\,GS/s.
Experimental conditions cube: setup 2, Broadcom, BC422 (cube, black, grease), rise time 0.77\,ns,
bandwidth 580\,MHz, sampling rate 20\,GS/s.
}
\label{Fig:stripe-sigma1MeV}
\end{figure}

Figure~\ref{Fig:stripe-sigma} shows the measured time resolution of the stripe detector as a function of the trigger fraction. The obtained value $\sigma \approx 20$\,ps is at least a factor of 2 below any result we ever measured in practice before. Time resolution per 1\,MeV detected energy extracted from the above data, is compared in Fig.\,\ref{Fig:stripe-sigma1MeV} with that measured with a BC422 painted black scintillator cube under the same experimental conditions. The observed degradation of $\sim 15$\,\% we attribute to an increased PTS contribution.

\section{Conclusion}

In \cite{Gundacker20} an exceptional time resolution of 5.3\,ps (sigma) per 1\,MeV detected energy was demonstrated with the BC422 scintillator. The result has been made possible through the use of low SPTR value SiPMs and a dedicated low-noise signal processing at 1.5\,GHz bandwidth. Also the rise time of BC422 scintillation was revisited and shown not to be a limiting factor for the time resolution of such detectors.

In this work we use our standard technique (signal processing bandwidth $\sim$\,600\,MHz, constant fraction discriminator) to measure the time resolution of BC422 read out by different SiPMs. 
The best results ${\rm CTR} = 12.5$\,ps (sigma) and $\sigma_{\rm 1MeV} = 5.9$\,ps$\times$MeV$^{0.5}$ obtained with a small ($3 \times 3 \times 3$)\,mm$^3$ scintillator are comparable with those reported in \cite{Gundacker20}.

Therefore we confirm that the time resolution with BC422 scintillator can be substantially improved by using low SPTR values SiPMs. 

\section{Acknowledgment}

We thank Florian Barchetti for cutting and polishing the scintillators, Andreas Hofer for his contribution in preparing the detector setup, Urs Greuter and Mario Liechti for the help in preparation of SiPMs in TSV packages for measurements. Discussions on several topics with Urs Greuter were very fruitful and are very much appreciated.

\end{document}